\documentclass[10pt, conference]{IEEEtran}

\usepackage{hyperref}
\usepackage{cite}
\usepackage{amsmath,amssymb,amsfonts}
\usepackage{algorithmic}
\usepackage{graphicx}
\usepackage{textcomp}
\usepackage{xcolor}
\usepackage{float}
\usepackage{stfloats}
\usepackage{listings}

\definecolor{ForestGreen}{RGB}{0, 155, 85}

\begin{document}

\title{Tydi-lang: A Language for Typed Streaming Hardware}

\DeclareRobustCommand*{\IEEEauthorrefmark}[1]{%
  \raisebox{0pt}[0pt][0pt]{\textsuperscript{\footnotesize #1}}%
}

\author{\IEEEauthorblockN{
Yongding Tian\IEEEauthorrefmark{1} \ \ \ \ \ 
Matthijs A. Reukers\IEEEauthorrefmark{1} \ \ \ \ \ 
Zaid Al-Ars\IEEEauthorrefmark{1} \ \ \ \ \ 
Peter Hofstee\IEEEauthorrefmark{1,}\IEEEauthorrefmark{2}}
\IEEEauthorblockN{
Matthijs Brobbel\IEEEauthorrefmark{3} \ \ \ \ \ 
Johan Peltenburg\IEEEauthorrefmark{3} \ \ \ \ \ 
Jeroen van Straten\IEEEauthorrefmark{3}
}
\IEEEauthorblockA{\IEEEauthorrefmark{1}Delft University of Technology, Delft, The Netherlands
}

\IEEEauthorblockA{\IEEEauthorrefmark{2}IBM Infrastructure, Austin, TX, US}

\IEEEauthorblockA{\IEEEauthorrefmark{3}Voltron Data, Mountain View, CA, US \\
Email: Y.Tian-3@tudelft.nl
}
}

\maketitle

\begin{abstract}


Transferring composite data structures with variable-length fields often requires designing unique protocols, causing incompatibility issues and decreased collaboration among hardware developers, especially in the open-source community. Because the high-level meaning of a protocol is often lost in translation to low-level languages when a custom protocol needs to be designed, extra documentation is required, the interpretation of which introduces new opportunities for errors. 



The Tydi specification (Tydi-spec) was proposed to address the issues by codifying the complex structures in a type and providing a standard protocol to transfer typed data among components. This paper presents Tydi-lang, a language that incorporates Tydi-spec for describing typed streams and offers templates for reusable components. An open-source compiler from Tydi-lang to Tydi intermediate representation (Tydi-IR) is implemented, and a Tydi-IR to VHDL compiler is utilized. Through Tydi-lang examples translating high-level SQL to VHDL, we demonstrate its efficiency in raising abstraction levels and reducing design effort.

\end{abstract}


\section{Introduction}

This paper proposes a high-level hardware description language called Tydi-lang, based on the type system introduced in the Tydi-spec \cite{tydi_spec}. 
The goal of Tydi-lang is to raise the abstraction level of typed streaming hardware and reduce the design effort for hardware designers. 
A typical compilation process for Tydi-lang, to VHDL for example, consists of two steps. The first step is compiling Tydi-lang to Tydi-IR\cite{tydi_ir} with the Tydi-lang compiler, which will be discussed in this paper. The second step, compiling Tydi-IR to VHDL, is implemented by the Tydi-IR project. Unlike Tydi-IR, Tydi-lang is designed for developers with the intent of reducing design effort and raising the level of abstraction.

Based on Tydi-spec and Tydi-IR, Tydi-lang introduces a generative syntax and a template concept, which allows developers to describe hardware components in a more abstract and reusable way. These two features also allow developers to design streaming hardware more efficiently by directly connecting components at a higher level and facilitate translating software languages to Tydi-lang. Some frequently used component templates are introduced in a standard library for Tydi-lang. One of the benefits of using the Tydi-lang standard library is that developers can design digital circuits without having to use low-level HDLs, for example, to accelerate SQL queries via FPGA accelerators, where operations on data can be mapped to hardware templates. 

Besides the standard library, the Tydi-lang also integrates a high-level design rule check system to identify type errors, which would be un-trackable on the lower layer. The ability to continue to do type-consistent analysis at the lower levels motivates us to pursue a simulator infrastructure specific to Tydi. We present an event-driven simulation syntax to describe the component behavior. With the future Tydi-lang simulator, we could simulate the whole circuit and generate Tydi-IR testbenches for other synthesis tools to ensure hardware correctness.

The contributions of this paper can be summarized as follows (please notice that "high-level" in this paper does not refer to "high level synthesis"):
\begin{itemize}
    \item Design a user-friendly, type-safe, and high-level HDL for streaming hardware and implement its compiler (Tydi-lang).
    \item Introduce the "template" concept for typed streaming hardware. The template concept can be applied to customize components, describe components with shared behaviors, and describe components whose behaviors are independent of types.
    \item Provide a new toolchain (Tydi tools) to design FPGA accelerators for big data applications efficiently. This use case might be a foundation for a future trans-compiler from software programming languages to hardware description languages.
    \item Present a high-level simulator blueprint to facilitate design analysis, including identifying streaming bottlenecks and generating testbenches for low-level verification tools. This simulator also allows splitting the hardware designing tasks into high-level and low-level tasks and allows designers to cooperate with other designers on different levels.
\end{itemize}

This paper focuses on the Tydi-lang features. The organization of the paper is as follows: Section \ref{section:background} presents pertinent background information. Section \ref{section:tydi_lang_work_flow} describes the workflow for using the Tydi-lang toolchain and the workflow to accelerate big data applications, starting from mapping memory data structure to hardware ports, and ending at VHDL generation. Section \ref{section:tydi_lang_features} describes the key features of Tydi-lang and how these features can impact the development of streaming hardware.
Section \ref{section:tydi_lang_simulator_and_testbench_generation} describes the blueprint of the Tydi-lang simulator, which could identify deadlock and bottlenecks for streaming hardware. This section also discusses a method for generating VHDL testbenches for components designed by other tools. Section \ref{section:result} uses Tydi-lang to rewrite some queries in the TPCH benchmark and shows how efficient it could be in designing FPGA accelerators. Section \ref{section:conclusion} concludes the paper.

\section{Background}
\label{section:background}
FPGAs have found use as accelerators in big data applications because the performance increase in general-purpose processors cannot keep up with the continued increase in big data sizes \cite{FPGA_Acceleration_for_Big_Data_Analytics_Challenges_and_Opportunities}. Unlike designing software, FPGA applications are still often designed with a register-transfer level (RTL) language, which is more complex than general software languages. In addition, designing FPGAs requires hardware specialists because the hardware workflow is longer and specific knowledge is required. Thus, many automation tools and frameworks have come out to assist in building FPGA applications. These tools can be categorized into two types: 1) generate IP cores from software languages and wrap FPGAs with nice software interfaces or 2) generate high-level synthesis (HLS) or RTL code automatically for domain-specific applications. Examples of the first type, some vendor tools provide tools to specify FPGA accelerators in OpenCL. This method can greatly reduce the line of code (LoC), but the generated circuits usually consume more FPGA resources (x3.56) and show worse performance (x0.56) \cite{A_Study_on_the_State_of_High_Level_Synthesis}. \cite{fpga_acceleration_rnn} and \cite{hls4ml} are good examples of the second type, as they generate FPGA code for machine learning, but the purpose of generated code is domain-specific and not flexible. Even though OpenCL provides a standard interface to write general-purpose acceleration algorithms on FPGAs, it does not focus on the problem of delivering data with complex structures from memory to FPGAs. Users usually need to serialize the memory data or manually design their own bit-level protocols to send memory data to FPGAs or send data from components to components. 

To address the issue of sending complex data among components and from memory to FPGAs, \cite{tydi_spec} proposed the Tydi-spec, which provides a standard way to map data structures to hardware streams. In Tydi-spec, all data structures are constructed using four logical types: \texttt{Bit}, \texttt{Group}, \texttt{Union} and \texttt{Stream}. \texttt{Bit} represents the hardware bit required to represent a value. For example, an ASCII character requires \texttt{Bit(8)} to represent. \texttt{Group} is a combination of other types, and the total bit width is the sum of all child-type bit widths. \texttt{Union} represents the data can be one of the types, and the largest child bit width determines the union bit width. \texttt{Stream} describes the stream-space properties of a logical type such as the dimension, direction of stream, throughput, etc. For example, \texttt{Stream(Bit(8), dimension=2)} can be used to represent an English sentence because each character is a \texttt{Bit(8)}. The word length and the number of words in this sentence are unknown, so the dimension is 2. \texttt{Stream} also defines the hardware protocol and the handshaking mechanism from source to sink.

Based on Tydi-spec, a language called Tydi-Intermediate Representation (Tydi-IR\cite{tydi_ir}) is proposed to describe the type system in Tydi-spec, with an extension of describing hardware components. In Tydi-IR, users can define the port map of a component as a \texttt{streamlet}, and each port must bind to a \texttt{stream} type. This type system ensures only two compatible ports can be connected. To describe the behavior of the component, developers can use \texttt{implementation} to describe its inner implementation instance and connections (\texttt{streamlet} is similar to \texttt{entity} and \texttt{implementation} is similar to \texttt{architecture} in VHDL). The process of compiling Tydi-IR to VHDL is discussed in a separate paper.

However, Tydi-IR, like many other intermediate representations, usually contains excessively precise details of hardware components. This is not suitable for developers because it usually takes too long to write. In addition, the lack of abstraction in Tydi-IR makes development onerous. For example, developers need to define many different stream duplicators for each logical type, though these duplicators are similar on the VHDL level. Therefore, this paper proposes Tydi-lang to reduce the design effort.

CHISEL \cite{chisel} is a hardware construction language embedded in Scala. It raises the level of abstraction in hardware design by using variables and flow control statements to construct hardware and by adopting object-oriented patterns in Scala. Extending CHISEL with Tydi type specifications seems an attractive solution that can directly take advantage of CHISEL and Scala. However, we decided to pursue Tydi-lang as a separate project because we want to design a concise prototype first, independent of any specific design language.


FPGAs can efficiently accelerate big data analytics applications by accessing the memory data via PCI-E or OpenCAPI and processing the data with customized circuits \cite{Zero_Cost_Arrow_Enabled_Data_Interface_for_Apache_Spark} \cite{Battling_the_CPU_Bottleneck_in_Apache_Parquet_to_Arrow_Conversion_Using_FPGA}. Fletcher \cite{fletcher} is a tool to automatically generate the hardware components for FPGA accelerators to access the Apache Arrow \cite{arrow} data stored on the host memory. The data structure is described by its Arrow schema. Tydi-lang can take advantage of Fletcher to map the Arrow data structures to Tydi-lang logical types and quickly use the components in Tydi-lang standard library to construct the FPGA accelerator.

\section{Tydi-lang workflow}
\label{section:tydi_lang_work_flow}

\begin{table*}[]
\centering
\caption{Terms used in Tydi-spec and Tydi-IR}
\label{table:tydi_ir_terms}
\resizebox{\textwidth}{!}{
\begin{tabular}{|c|c|l|}
\hline
Term & Type & \multicolumn{1}{c|}{Meaning} \\ \hline
Null & Logical type & Represents empty data. A stream of null type will be optimized out. \\ \hline
Bit(x) & Logical type & Represents data that requires x hardware bits to represent. \\ \hline
Group(x,y) & Logical type & \begin{tabular}[c]{@{}l@{}}A tuple of several other logical types (x and y in this example). The number of hardware bit \\ would be the sum of all child element bit width.\end{tabular} \\ \hline
Union(x,y) & Logical type & \begin{tabular}[c]{@{}l@{}}An union of several other logical types (x and y in this example). The number of hardware bit\\ would be the maximum bit width of a single child.\end{tabular} \\ \hline
Stream(x) & Logical type & \begin{tabular}[c]{@{}l@{}}Represents a stream of a logical type. The stream can also specify the data dimension,\\ protocol complexity, hardware synchronicity, and throughput as optional arguments.\end{tabular} \\ \hline
Port & Hardware element & Represents a hardware port, the port must specifies its logical stream type and direction. \\ \hline
Streamlet & Hardware element & \begin{tabular}[c]{@{}l@{}}Represents the port map of a component. This term is almost the same as the "entity" \\ term in VHDL.\end{tabular} \\ \hline
Implementation & Hardware element & \begin{tabular}[c]{@{}l@{}}Represents the inner structure of a component. The inner structure should be a combin-\\ ation of instances and connections. Implementation must specify a streamlet as its port \\ map, this relationship is similar to the relationship between "entity" and "architecture" in \\ VHDL. Implementation can be declared as "external" if they cannot be represented by \\ instances and connections. "Implementation" is abbreviated to  "impl" in Tydi-lang. \end{tabular} \\ \hline
Connection & Hardware element & \begin{tabular}[c]{@{}l@{}}Connect two ports. The two ports must have the same data stream type, compatible \\ protocol complexities, correct directions and same clock domain. Connections must be \\ declared in implementation.\end{tabular} \\ \hline
Instance & Hardware element & \begin{tabular}[c]{@{}l@{}}Represents a nested implementation instance in another implementation. The port of the \\ nested  implementation can be accessed by using the instance.\end{tabular} \\ \hline
Clock domain & Hardware Clock & \begin{tabular}[c]{@{}l@{}}A clock domain is a representation of clock frequency and phase and is usually bound to \\ a port. Due to the handshaking mechanism in the stream, the clock domain concept \\ ensures only two ports with the same clock domains can be connected together. \end{tabular} \\ \hline
\end{tabular}%
}
\end{table*}

\begin{figure}
    \centering
     \includegraphics[width=0.95\columnwidth]{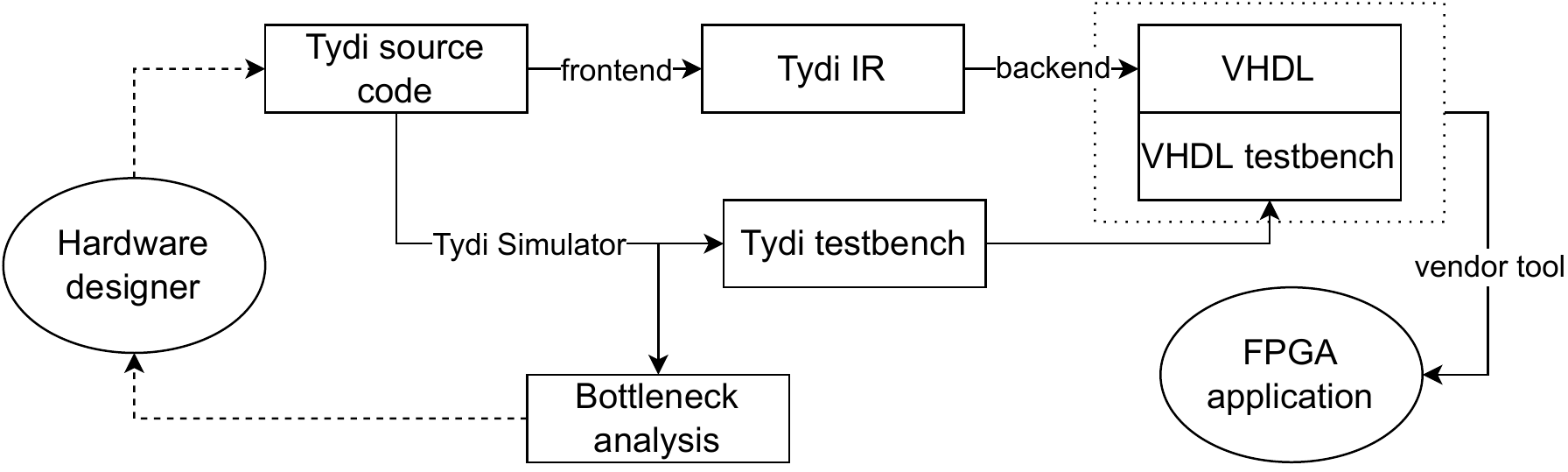}
    \caption{Tydi-lang toolchain workflow}
    \label{fig:tydi_lang_tool_chain_work_flow}
\end{figure}

Table \ref{table:tydi_ir_terms} lists the terms and concepts used in Tydi-spec\cite{tydi_spec} and Tydi-IR that appear in this paper.
The Tydi toolchain workflow is provided in Figure \ref{fig:tydi_lang_tool_chain_work_flow}. The flow starts with hardware designers writing Tydi-lang source code to describe the streaming hardware. As mentioned in Table \ref{table:tydi_ir_terms}, there are two types of implementations. For external implementations, designers can write "simulation code" to describe the behavior. The "simulation code" can be translated to Tydi-IR testbenches by the Tydi simulator and further converted to a VHDL testbench by the Tydi-IR tool. The VHDL testbench is important for hardware verification because external implementations are designed by external tools where the testbench can ensure the actual behavior matches the Tydi side. Another function of the Tydi simulator is analyzing the bottleneck of streaming hardware and identifying potential deadlocks if the simulation of all used external implementations are defined.

The Tydi-lang source code is compiled to Tydi-IR by the Tydi frontend and is further compiled (in our toolchain to VHDL) by a Tydi backend. The output VHDL code and testbench, with external VHDL code, can together be synthesized to generate a bitstream file for an FPGA.

\begin{figure}
    \centering
  \includegraphics[width=0.95   \columnwidth]{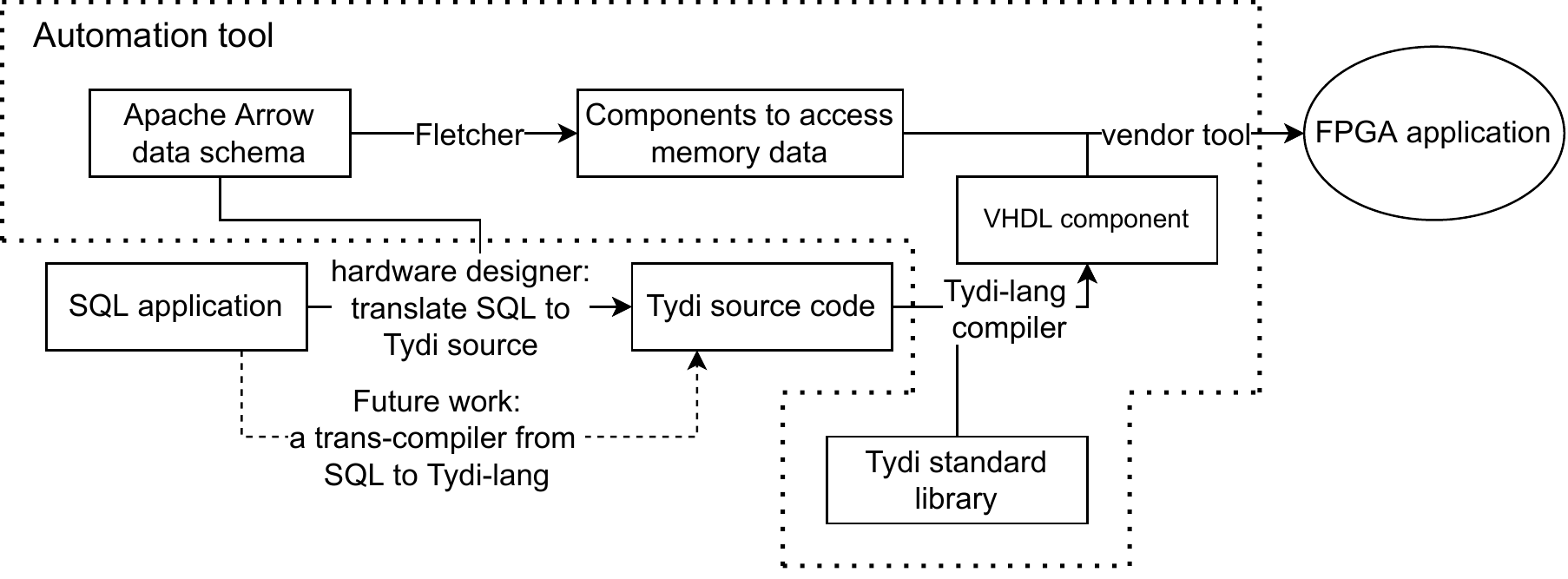}
    \caption{Tydi-lang workflow in big data}
    \label{fig:tydi_lang_in_big_data}
\end{figure}

To illustrate the use of Tydi we provide an example from big data, Tydi-lang can be an elegant bridge to connect query languages and the FPGA accelerators, as shown in Figure~\ref{fig:tydi_lang_in_big_data}. Big data developers usually use SQL to do analytics on a dataset with a known schema. We use Apache Arrow as the dataset format because it is widely applied in big data applications for zero serialization overhead. With Fletcher \cite{fletcher}, which is a tool to generate hardware components to access Apache Arrow data automatically, the design effort can be greatly reduced while the only thing left to do is translating the SQL to Tydi-lang. Our experience suggests it is possible to design a tool to automatically compile SQL to Tydi-lang in the future.

\begin{figure}
    \centering
     \includegraphics[width=0.95\columnwidth]{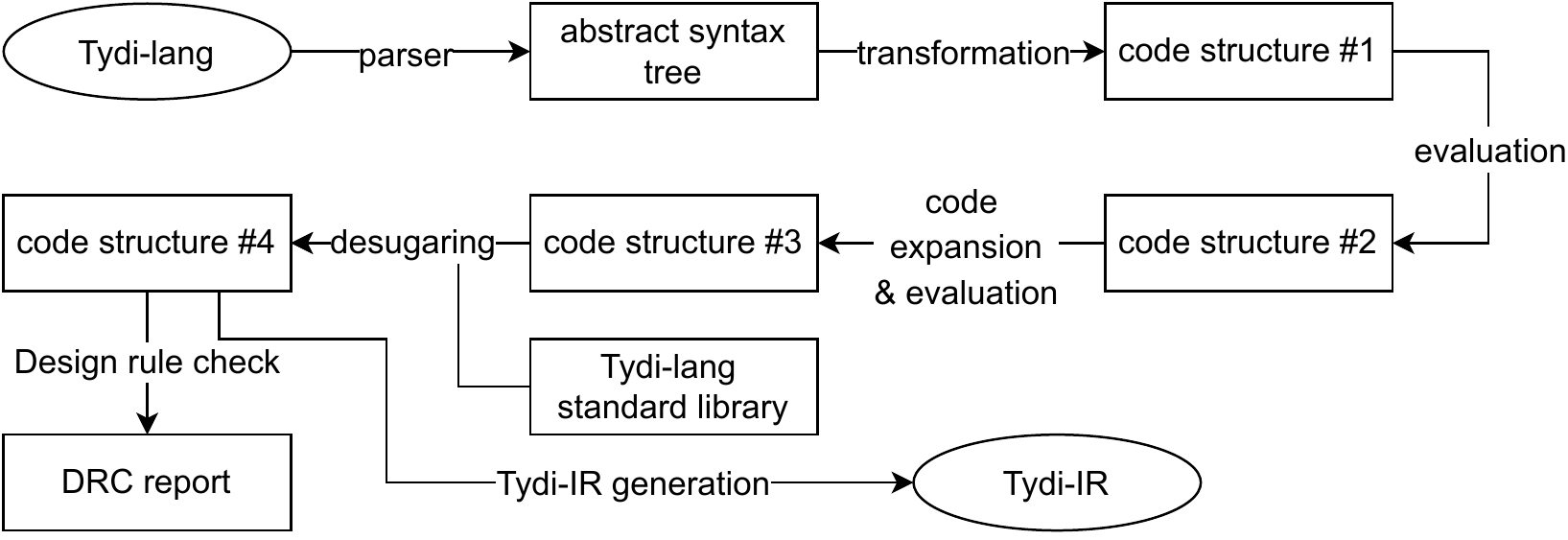}
    \caption{Workflow of Tydi-lang compiler frontend}
    \label{fig:tydi_front_end}
\end{figure}

As mentioned, this paper focuses on the frontend of the Tydi-lang compiler. A detailed illustration of the frontend workflow is provided in Figure \ref{fig:tydi_front_end} and discussed in~\cite{tian2022tydilang}. 
The term "code structure" denotes an intermediate representation of the source code, and the number indicates the versions. The design rule check (DRC) will check two rules. The first rule checks that the logical types of two connected ports must be identical to avoid misinterpreted data. This rule check is essential because the type information will be removed after generating VHDL. The second rule checks the port usage count because each port must be used once under the handshaking mechanism.

\section{Tydi-lang features}
\label{section:tydi_lang_features}
This section discusses some language-level features of Tydi-lang and how these features assist developers in designing streaming hardware.

\subsection{Hardware description by variables}
\label{section:hardware_description_by_variables}
There are five types of variables in Tydi-lang: integer, floating-point number, string, boolean, and clock domain. The clock domain type is the same as that in Table \ref{table:tydi_ir_terms} and the other four types are the same as those in general software programming languages. In Tydi-lang, all variables must be immutable because the values of mutable variables depend on the execution flow, which does not exist in hardware. Variable shadowing is possible and useful if users want to present extra information in a smaller scope. For an example of applying a variable, in SQL, \texttt{decimal(10,2)} is equivalent to \texttt{decimal(10)} on the hardware layer, and developers might not distinguish them. In Tydi-lang, they can be easily differentiated by defining two logical group types which contain an integer separately to indicate the number of digits after the digit point.

Tydi-lang also provides a math system to calculate the variable values by expression. This math system can help developers to write more flexible code than VHDL. For example, developers need to write a function in VHDL to calculate the bit width required to represent \texttt{Decimal(15)} in SQL to calculate. But in Tydi-lang, the expression 

\texttt{ Bit(ceil(log2(10$\wedge$15-1)))}
can directly represent the bit width as a logical bit type. The number "15" can be further replaced by a variable to get better flexibility (e.g. \texttt{Bit(ceil(log2(10$\wedge$ decimal\_width\_memory-1)))}).

In addition to supporting basic variables, Tydi-lang also provides "array" to contain multiple values. A single array variable can be expanded to multiple hardware components with generative "for" syntax. This "array" concept is helpful in translating memory arrays to hardware designs. Consider following SQL statement: \texttt{where\ p\_container\ in\ ('MED\ BAG', 'MED\ BOX', 'MED\ PKG', 'MED\ PACK') }, which filters based on \texttt{p\_container}. There should be four comparators and a 4-port logical "or" component to achieve the filtering. In Tydi-lang, the four strings can be stored in a string array and then use the "for" statement to declare four instances of a comparator template that receives a string as the standard input. As for the "or" component, we can define an implementation template that receives an integer as input count. Then the input of "or" can be defined as a port array, and developers can use single for statement to connect each comparator and input of "or" (e.g. \texttt{for i in (0-1->count)\{ comparator[i].output => or.input[i];\}} ). This syntax is useful for presenting memory arrays with elements or sizes that are not known when writing code. Tydi-lang also provides "if" syntax to control whether to generate the implementation instances and connections inside the "if" scope. The "if" syntax can handle type-specific and component-specific cases. For example, an "adder" template works for both 32-bit integer values and \texttt{decimal(32bit, 2)} values because the adders for the two types are the same on hardware. But for a "multiplier" template, they are different because the output of the decimal input should be \texttt{decimal(32bit, 4)}. Here template writers can use "if" and "assert" to restrict the logical type to avoid potential errors.

The features such as \texttt{if}, \texttt{for} and \texttt{assert} are summarized in Table \ref{table:tydi_var_based_features}.

\begin{table*}[]
\centering
\caption{Features based on variables in Tydi-lang}
\label{table:tydi_var_based_features}
\resizebox{\textwidth}{!}{
\begin{tabular}{|c|c|l|}
\hline
Term & Type & \multicolumn{1}{c|}{Meaning} \\ \hline
\begin{tabular}[c]{@{}c@{}}for x in x\_array \\ \{  /*scope*/ \}\end{tabular} & syntax & \begin{tabular}[c]{@{}l@{}}the Instances and connections in the scope will be expanded to the external scope with each\\ value x in x\_array. x\_array must be an array of basic values.\end{tabular} \\ \hline
\begin{tabular}[c]{@{}c@{}}if (x) \\ \{ /*scope*/ \}\end{tabular} & syntax & \begin{tabular}[c]{@{}l@{}}x must be a boolean value, the instances and connections will be expanded to the external\\ scope if x is true.\end{tabular} \\ \hline
assert(var) & builtin function & Assert the variable var is true. Throw an error if it is false. \\ \hline
\end{tabular}%
}
\end{table*}

\subsection{Abstract hardware template}
\label{section:abstract_hardware_template}
The "template" concept in Tydi-lang is similar to the "class template" in c$++$. More specifically, an abstract hardware template describes a set of different components to process input streams with shared properties (e.g., adders for integer and decimal), or components that are independent of data types (e.g., handshake protocol level hardware), or components which could be configured to act differently (e.g., a configurable constant integer generator). A template will never be compiled to Tydi-IR, and only template instances, where the template arguments are provided, will be compiled to Tydi-IR.

In Tydi-lang, only streamlets and implementations can be declared as templates. It is also reasonable to have logical type templates(e.g., a group type containing a child type with unknown bit width), but we choose not to do so because it introduces the type equality problem when declaring connections. The type equality problem is that developers might design two types with the same number of hardware bits, but they are for different purposes and should not be connected. A real case could be two type template instances that are instantiated with the same variable value but with different variable names. It is hard to determine whether the two types are equal or not. In addition, DRC will check the strict type equality (two ports must be defined with the same logical type variable) for all connections by default. Adding an extra attribute can disable the strict type equality checking and turn to verify the equality of type hierarchy.

As for the template arguments, previous examples have already shown that variables and logical types can be template arguments. Besides these two, a streamlet can also be a template argument. However, the streamlet template only accepts implementations derived from that streamlet when the template is instantiated. This mechanism is helpful if developers want to design a component with unknown components inside. For example, a developer would like to design a component to fully utilize the bandwidth by parallelizing the processing unit with a data de-multiplexer and a data multiplexer. The streamlet template can be an abstract representation of that processing unit with a known interface but unknown implementation. The source Tydi-lang code is listed below.

In addition to template arguments, defining a template argument is equivalent to defining the corresponding variable, type, or implementation in the template scope. This feature also allows developers to pass a template argument to another template, which is a frequently used pattern in Tydi-lang. For example, this code snippet, "\texttt{impl\ void\_i<type\_in:\ type>\ of\ void\_s<type\ type\_in>}", shows passing template type "type\_in" from an implementation to a streamlet.

 \begin{lstlisting}
//below is a streamlet template with 2 template arguments
//data_type: a logic type representing the input/output logic stream type
//channel: number of processing unit
streamlet parallelize_s<in_data_type: type, out_data_type: type> {
  input: input_data_type in,
  output: output_data_type out,
}

/*process_unit: process the data packet*/
streamlet process_unit_s<input_data_type: type, output_data_type: type> {
  input: input_data_type in,
  output: output_data_type out,
}

/*the implementation of parallelize_i component*/
impl parallelize_i<in_data_type: type, 
    out_data_type: type, 
    //the processing unit should be an  
    //implementation of process_unit_s
    pu_instance: impl of process_unit_s, 
    channel: int> 
    of 
    parallelize_s<type in_data_type, 
    type out_data_type> 
    {
    //a demux to transfer the data packets to 
    //different processing units
    instance demux_inst(demux_i<type input_data_type, channel>),
    //a mux to select the proper output
    instance mux_inst(mux_i<type out_data_type, channel>),
    //processing units
    instance pu(pu_instance) [channel],
    //connect the mux/demux and process units
    for i in 0-1->channel {
        demux_inst.out[i] => pu[i].in,
        pu[i].out => mux_inst.in[i],
    }
}
\end{lstlisting}

Let's assume we possess a 32-bit adder with a delay of 8 clock cycles. In order to process an input stream at a rate of 1 data packet per clock cycle, we can employ the \texttt{parallelize} component as outlined below, ensuring we achieve the desired throughput.

\begin{lstlisting}
//adder input data stream
Group AdderInput {
    data0: Bit(32),
    data1: BIt(32),
}
type Input = Stream(AdderInput);

//result stream
Group Bit32_result {
    data: Bit(32),
    overflow: BIt(1),
}
type Result = Stream(Bit32_result);

//the implementation of adder_32 goes here
impl adder_32 of process_unit_s<type AdderInput, type Bit32_result> {
    ...
}

//achieving 1 data/cycle
instance parallelize_i<type Input, type Result, impl adder_32, 8>;
\end{lstlisting}

The template concept, with the "hardware description by variable" concept in Section \ref{section:hardware_description_by_variables}, is the most important method in describing abstract components. Using the abstract components is encouraged in Tydi-lang projects because they can greatly improve code reuse-ability and flexibility. We summarize some frequently-used abstract components available in the Tydi-lang standard library.

\subsection{Tydi-lang standard library}
\label{section:tydi_lang_standard_library}

The Tydi-lang standard library is a pure-template library, defining many frequently utilized components that can be categorized into the following three types.

\begin{itemize}
    \item Components to duplicate/remove data packets. The Tydi-lang is designed for streaming hardware where each port can only be connected once, while using a value several times is common in software languages. Thus duplicator and voider (a component name) are proposed to duplicate data packets and remove data packets. In the low-level implementation, duplicators copy and resend the bit-level data to all output ports and only acknowledge the input port when all outputs are acknowledged. Voiders will remove all data packets by always acknowledging the source component and ignoring the data. These two components work on the handshaking layer and hardware bit, so they are templates in Tydi-lang. 
    \item Components that describe common behaviors for different logical types. For example, an adder can work for integer types, decimal types, and many other numerical types once the bit width is specified. A comparator is also possible to compare integers, dates, etc. However, selecting and implementing these components might be tricky because the multipliers for integer and decimal are different (if taking the digits after the digit point into consideration). For this case, assertion and "if" can be applied to restrict the template.
    \item Components to transform logical types. The transformation includes splitting a group type into its inner types or combining several logical types in a group. These template components help process ports with user-defined composite data structures. This part is future work and has not been implemented in the current Tydi-lang version.
\end{itemize}

Unlike typical template components, the components in the Tydi-lang standard library are too elementary to be described as instances and connections (external implementations if using terms in Table \ref{table:tydi_ir_terms}), so there is another RTL generation process for these standard components. However, this generation process must be manually defined. For example, in a duplicator template with two arguments - a logical type of stream and an integer variable to indicate the output port count, the process to generate the correct component needs to be hardcoded into the generator.

Because adding a new component template in the Tydi standard library means adding more hard-coded processes in the generator, the Tydi-lang standard library should be kept as small and as abstract as possible. It is a compromise between the library size and the generator complexity, resulting in greater difficulty in designing the standard library. In addition, finding the proper abstraction of each component is also complicated. The selection of components in the Tydi-lang library and their corresponding templates remain under construction. The Tydi-lang library used in Section \ref{section:result} is a prototype and only includes the essential templates for our test cases. 

\subsection{Sugaring the hardware design}
\label{section:sugaring_the_hardware_design}
\begin{figure}
    \centering
     \includegraphics[width=0.98\columnwidth]{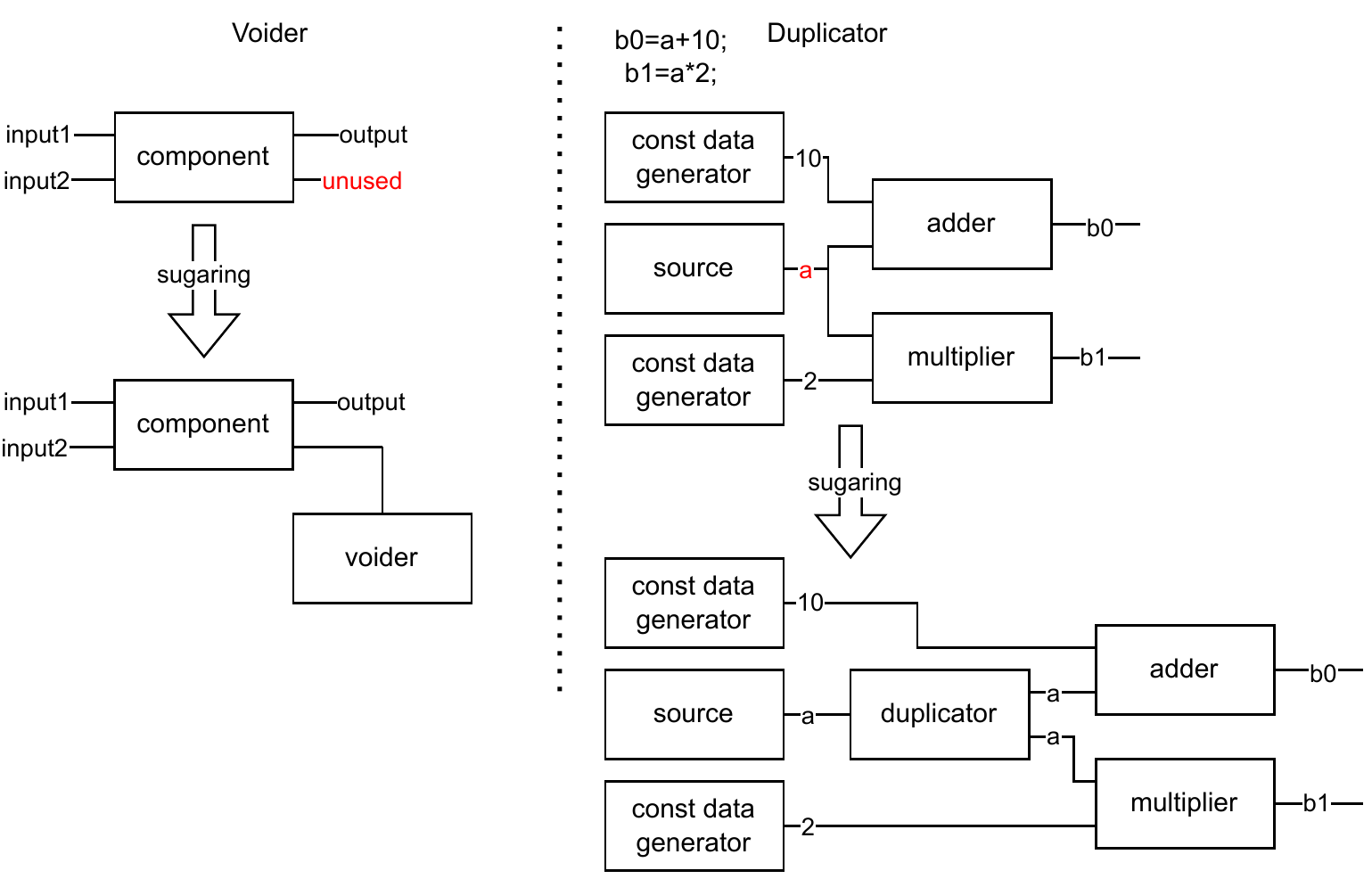}
    \caption{Auto insertion of voider and duplicator}
    \label{fig:sugaring}
\end{figure}

Sugaring is an important factor in reducing language developers' design effort by automatically inferring and appending the absent code. With the help of the Tydi-lang standard library, the current compiler provides two types of sugaring. The first type of sugaring is the automatic duplicator template insertion if an output port has been connected to multiple input ports. The compiler will automatically infer the logical type and the output channel size of the duplicator template. The second type of sugaring is the automatic voider template insertion if an output port has never been used, where voider is a component that does nothing but is always ready to receive the next packet. These two sugarings release the restriction that "one port must be connected to exactly one other port", as illustrated in Figure \ref{fig:sugaring}.

For sugaring examples, consider the case of using Fletcher \cite{fletcher_internals} to generate components to access memory data from a data schema. The data schema might be large while the query only accesses a small portion of it, and the query on data is flexible while the generated components are rigid. Without sugaring, developers need to manually append voiders for each unused port on the generated Fletcher components. 

\subsection{Comparison}
\label{section:comparision}
Table \ref{table:comparison} compares the Tydi-lang with other popular HDLs based on a survey \cite{CHISEL_survey} that summarizes many existing high-level HDLs. The criteria "supports design aspects" includes three aspects: 1) architecture: describe the connections and components, 2) configuration: customize components, and 3) functionality: describe the component behaviors. The FP and OOP in "Paradigm support" stand for functional programming and object-oriented programming.

\begin{table}[]
\centering
\caption{Compare Tydi-lang with other high-level HDLs}
\label{table:comparison}
\resizebox{\columnwidth}{!}{
\begin{tabular}{|l|l|l|l|l|}
\hline
Languages & Base language & \begin{tabular}[c]{@{}l@{}}Supported design \\ aspects\end{tabular} & Paradigm support & Output \\ \hline
Genesis2\cite{Genesis2} & SystemVerilog & \begin{tabular}[c]{@{}l@{}}Architecture \\ configuration \\ functionality\end{tabular} & OOP & HDL \\ \hline
Clash\cite{clash} & Haskell & \begin{tabular}[c]{@{}l@{}}Architecture \\ configuration \\ functionality\end{tabular} & FP & HDL \\ \hline
Vitis HLS\cite{A_Study_on_the_State_of_High_Level_Synthesis} & \begin{tabular}[c]{@{}l@{}}Most prominently \\ C/C++-based\end{tabular} & \begin{tabular}[c]{@{}l@{}}Architecture \\ configuration \\ functionality\end{tabular} & \begin{tabular}[c]{@{}l@{}} bit-level stream,\\ FP, OOP with \\ templates \end{tabular} & HDL \\ \hline
CHISEL\cite{chisel} & Scala & \begin{tabular}[c]{@{}l@{}}Architecture \\ configuration \\ functionality\end{tabular} & \begin{tabular}[c]{@{}l@{}} bit-level stream, \\ FP, OOP with \\ templates \end{tabular}  & HDL, FIRRTL \\ \hline
Kamel \cite{kamel} & IP-XACT & Architecture & other & HDL \\ \hline
Veriscala \cite{Veriscala} & Scala & \begin{tabular}[c]{@{}l@{}}Architecture \\ configuration \\ functionality\end{tabular} & FP, OOP & HDL + driver (FPGA) \\ \hline
Tydi-lang & \begin{tabular}[c]{@{}l@{}}None \end{tabular} & \begin{tabular}[c]{@{}l@{}}Architecture\\ configuration\end{tabular} & \begin{tabular}[c]{@{}l@{}}built-in typed \\ stream, OOP \\ with templates\end{tabular} & \begin{tabular}[c]{@{}l@{}}Depends on the\\ Tydi-IR backend,\\ currently supports \\ VHDL\end{tabular} \\ \hline
\end{tabular}
}
\end{table}

Compared with other high-level HDLs, the unique features of Tydi-lang are the ability to encode the structured data with logic types and the abstraction mechanisms for the logic type system. Tydi-lang makes it possible to map software data types to hardware logic types directly and efficiently design the FPGA accelerator. Tydi-lang does not support describing functionality (behavior) because it is designed for high-level designers as discussed in Section \ref{section:tydi_lang_simulator_and_testbench_generation}.

\section{Tydi-lang simulator and testbench generation}
\label{section:tydi_lang_simulator_and_testbench_generation}
The goal of the Tydi-lang simulator is assisting high-level developers in designing streaming circuits to meet functional requirements regardless of low-level behavior, and generating testbenches to collaborate with low-level developers. 

Simulating the streaming hardware on the Tydi-lang level is necessary because the response time of a single component is determined by the arrival time of asynchronous input data packets. Analyzing the timing information of all components can quickly help designers identify streaming bottlenecks. Using traditional low-level simulators for such work is cumbersome because there are too many trivial low-level signals such as handshaking. Our simulator can also predict the output sequences under certain input sequences, but this is also possible with traditional simulation tools, so we will not address this feature in this paper.

This section will present three aspects of the Tydi-lang simulator currently under development: simulation syntax, and simulator mechanism. 

\subsection{Simulation syntax}
The Tydi-lang simulation code is defined inside an implementation to describe its behavior. A sample simulation code is available here \cite{simulation_syntax}. Implementation defined by inner instances and connections should not have simulation code because inner instances characterize its behavior. The simulation syntax includes the following parts.

\begin{itemize}
    \item State variable: represents a state with a string value.
    \item Acknowledge mechanism: because the Tydi-lang integrates the handshaking mechanism from Tydi-spec, it is crucial to control the handshaking behavior and time. For example, a component with two input ports with different throughputs should have synchronization on its ports. This synchronization can be achieved by controlling the time of acknowledging the output ports.
    \item Event-driven: an event is an action from ports, such as receiving a data packet. Designers can use boolean logic to define composite events. For example, only compute when both data from two ports are ready. The process when an event happens is called an event handler, where behavior code, such as sending acknowledge signals, changing state variables, sending data to other components, and delaying for a specific time, can be defined here. In addition, the "if" and "for" syntax is available in the event handler as logic flow control syntax.
\end{itemize}


\subsection{Simulator}

Performing simulation requires the input data sequence to top-level implementation and the mapping from the clockdomain to physical frequency and phase. The simulator can calculate the delay time, record data flows, and record the state-transition table of each implementation.

The delay time includes the delay from components simulation code and connection. The time to transfer data packets via connections is calculated with the connection clockdomain and data packet length. The data flow and the state transformation can be inferred from the simulation code. The state means the combination of all possible values of all state variables. Notice that some hardware components cannot be described by the "state" system, for example, the random number generator. 

Because state transformation is caused by events, which are combinations of receiving data from different ports, analyzing the relationship between data flow and state could also help identify the potential for deadlock. As for identifying bottlenecks, the simulator should be able to record the waiting time of all output ports (blocked by handshaking). Designers can investigate the output ports with the longest blockage to find the bottleneck component.

\subsection{Generate testbench}

While the simulation code only describes the expected behavior of components, it does not guarantee low-level behavioral correctness. The Tydi-lang simulator should be able to generate testbench files to ensure the expected behavior matches the low-level simulation results. Tydi-IR already defined a testbench syntax based on prediction strategy (giving certain input and verifying output correctness), and provided a tool to translate from Tydi-IR testbenches to VHDL testbenches. The Tydi-lang simulator can utilize this tool to generate VHDL testbenches.

The mechanism to generate testbench can be briefly described as an "input - current state - output" testing system. The "input" corresponds to an event in Tydi-lang, the "current state" is a combination of events and the initial state, and the "output" corresponds to sending data. Generating testbenches is a process of using the above mechanism to cover all states and events in the state transition table. The coverage of input data in the simulation stage is important because uncovered input results in uncovered state transformation. The testbench system also reduces design effort because only low-level components require simulation code and testbenches, which is easier than writing testbenches for high-level components.


\section{SQL to Tydi-lang cases}
\label{section:result}

\begin{table*}[htbp]
\centering
\caption{LoC for translating TPC-H queries to Tydi-lang}
\label{table:result}
\resizebox{\textwidth}{!}{
\begin{tabular}{|lc|c|c|cc|c|}
\hline
\multicolumn{2}{|r|}{LoC for Fletcher part ($LoC_f$)} & \multicolumn{1}{l|}{166} & \multicolumn{1}{l|}{} & \multicolumn{2}{r|}{LoC for Tydi-lang standard library ($LoC_s$)} & \multicolumn{1}{l|}{151} \\ \hline \hline

\multicolumn{1}{|c|}{Query name} & Raw SQL query & \begin{tabular}[c]{@{}c@{}}Query logic \\ in Tydi-lang \\ ($LoC_q$) \end{tabular} & \begin{tabular}[c]{@{}c@{}}Total Tydi-lang LoC \\ ($LoC_a$) \end{tabular} & \multicolumn{1}{c|}{\begin{tabular}[c]{@{}c@{}}Generated VHDL\\ ($LoC_{vhdl}$)\end{tabular}} & \begin{tabular}[c]{@{}c@{}}Ratio: \\ VHDL/Query logic\\ ($R_q$)\end{tabular} & \begin{tabular}[c]{@{}c@{}}Ratio:\\ VHDL/Total Tydi-lang\\ ($R_a$)\end{tabular} \\ \hline
\multicolumn{1}{|l|}{TPC-H 1 (without sugaring)} & 20 & 402 & 709 & \multicolumn{1}{c|}{7547} & 18.77 & 10.50 \\ \hline
\multicolumn{1}{|l|}{TPC-H 1} & 20 & 284 & 601 & \multicolumn{1}{c|}{7547} & 26.57 & 12.56 \\ \hline
\multicolumn{1}{|l|}{TPC-H 3} & 22 & 166 & 483 & \multicolumn{1}{c|}{6291} & 37.90 & 13.02 \\ \hline
\multicolumn{1}{|l|}{TPC-H 5} & 24 & 197 & 514 & \multicolumn{1}{c|}{6992} & 35.49 & 13.60 \\ \hline
\multicolumn{1}{|l|}{TPC-H 6} & 9 & 108 & 425 & \multicolumn{1}{c|}{4586} & 42.46 & 10.79 \\ \hline
\multicolumn{1}{|l|}{TPC-H 19} & 35 & 297 & 614 & \multicolumn{1}{c|}{11734} & 39.51 & 19.11 \\ \hline
\end{tabular}%
}
\vspace{-0.4cm}
\end{table*}

This section provides a use case of applying FPGAs to accelerate SQL queries to demonstrate the increased hardware abstraction level and the decrease in design effort. We translated several TPC-H SQL benchmark queries to Tydi-lang to represent the query logic on hardware and compare the line of code (LoC) of Tydi-lang and the generated VHDL. 

The implementation of the Tydi-lang compiler is written in Rust and open-sourced at the following link: \url{https://github.com/twoentartian/tydi-lang}. The implementation includes a parser\footnote{the parser grammar file is available in \url{https://github.com/twoentartian/tydi-lang/blob/main/tydi_lang_parser/src/tydi_lang_syntax.pest}} based on Rust-Pest~\cite{pest} and a number of transformation steps to transform Pest output to Tydi-IR (as shown in \autoref{fig:tydi_front_end}). Cloning the tydi-lang repository and compiling the "tydi\_compiler" crate should produce the executable compiler application. The "Cookbook" folder provides tutorials for each Tydi-lang feature and the Tydi-lang source code of each TPC-H query.

Here we take TPC-H query 19 (TPC-H 19) to illustrate how to convert a SQL query to Tydi-lang. The explanations are available in the SQL comments.

\begin{lstlisting}[
           language=SQL,
           basicstyle=\ttfamily\scriptsize, % font style
           showspaces=false,
           numbers=left,
           numberstyle=\tiny,
           commentstyle=\color{gray}
        ]
select
-- In tydi-lang, the 'select' clause is equivalent to component select<in: data_type, out: data_type, select_or_not: bool> where bool = Stream(Bit(1), d=1)
	sum(l_extendedprice* (1 - l_discount)) as revenue -- some arithmetic operations, can be abstracted to tydi templates of addition<in0: data_type, in1: data_type, out: data_type, overflow: bool>, multiply<...>...
from
-- Here are data tables that we need to get from Fletcher
	lineitem,
	part
where
-- we need a template component called filter<in: data_type, out: data_type, keep: Stream(bool)> here to remove unmatched data. The 'filter' component removes the current packet if the 'keep' signal is 0.
	(
        -- a combination of arithmetic operations
		p_partkey = l_partkey
		and p_brand = ':1'
		and p_container in ('SM CASE', 'SM BOX', 'SM PACK', 'SM PKG')
		and l_quantity >= :4 and l_quantity <= :4 + 10
		and p_size between 1 and 5
		and l_shipmode in ('AIR', 'AIR REG')
		and l_shipinstruct = 'DELIVER IN PERSON'
	)
	or (...)
	or (...)
        ;
\end{lstlisting}

As mentioned in \autoref{fig:tydi_lang_in_big_data}, each Tydi-lang project has three parts: the component generated by Fletcher to access memory, the query logic, and the standard library. We count the LoC of each part separately because the Fletcher part and standard library should not be counted in the design effort. We manually write the interface for Fletcher components because the current Fletcher project has not integrated Tydi-lang support yet. The primary keys in the TPC-H dataframe will be treated as input ports, and the other ports will be treated as output ports.

In addition, we also provide a non-sugaring version of the first query in TPC-H to show the design effort saved by sugaring. The results are provided in Table \ref{table:result}. The following formula calculates the calculation of ratio and total LoC:

\[ LoC_a = LoC_q + LoC_f + LoC_s \]
\[ R_q = LoC_{vhdl} / LoC_q \]
\[ R_a = LoC_{vhdl} / LoC_a \]

Other queries in the TPC-H benchmark are not tested in this paper because of the nested "select" structure in SQL. This structure requires storing an intermediate result back in memory for later processes, which is beyond our scope of testing the Tydi-lang compiler.

The result shows that designers can greatly reduce the line of code to design FPGA accelerators with the Tydi-lang. Because the code generated by Fletcher and the code in the Tydi-lang standard library should not be counted, the total design effort can be saved for over 40x compared with VHDL ($R_q$). For real cases, $R_q$ is determined by the level of abstraction and the query structure. Components with higher abstraction get higher $R_q$ because of templates. Queries that contain similar sub-structure can also get higher $R_q$. For example, the TPC-H 19 \cite{tpch_19} contains three "or" clauses with similar structure.

Please notice that the generated VHDL only includes hardware structure, and the current Tydi-lang compiler only generates hardware structure, too. In the future versions with a finished RTL generator for the standard library (mentioned in Section \ref{section:tydi_lang_standard_library}), the generated VHDL will contain behavior code, and the real ratio will be higher than the current result. 

\section{Conclusion and future work}
\label{section:conclusion}
This paper presents a new language (Tydi-lang) based on Tydi-spec to allow developers more effectively design streaming hardware. This new language also introduces the template concept to typed hardware, which raises the level of abstraction, saving design efforts for developers and enabling the possibility of translating software domain languages to hardware description languages. Along with the high-level language, we also present some verification and simulation tools to show how Tydi-lang works with low-level languages. We implement the Tydi-lang compiler prototype with its standard library and use several SQL query cases to demonstrate the new design flow and its effectiveness.

For future work, we intend to focus on using Tydi-lang in the big data analytics area by implementing the behavior part of the Tydi-lang standard library and improving Fletcher to generate Tydi interfaces. After that, we plan to integrate Tydi with other languages such as CHISEL, and develop more compiler backends for other low-level HDLs.

\section*{Acknowledgment}
This research was performed with the support of the Eureka Xecs project TASTI (grant no.~2022005).

\bibliographystyle{IEEEtran}
\bibliography{bibtex}

\end{document}